\documentclass[physrev,reprint,amsmath,amssymb,superscriptaddress,longbibliography]{revtex4-2}

\usepackage{graphicx}
\usepackage{dcolumn}
\usepackage{bm}
\usepackage{hyperref}
\usepackage{breakurl}
\usepackage{multirow}
\usepackage{enumitem}
\usepackage{color}

\begin{document}

\title{Metamagnetic Transitions in Few-Layer CrOCl Controlled by Magnetic Anisotropy Flipping}


\author{Minjie Zhang}
\thanks{Equal contributions}
\affiliation{Zhejiang Province Key Laboratory of Quantum Technology and Device, Department of Physics, and State Key Laboratory of Silicon Materials, Zhejiang University, Hangzhou 310027, P. R. China}

\author{Qifeng Hu}
\thanks{Equal contributions}
\affiliation{Zhejiang Province Key Laboratory of Quantum Technology and Device, Department of Physics, and State Key Laboratory of Silicon Materials, Zhejiang University, Hangzhou 310027, P. R. China}

\author{Chenqiang Hua}
\thanks{Equal contributions}
\affiliation{Zhejiang Province Key Laboratory of Quantum Technology and Device, Department of Physics, and State Key Laboratory of Silicon Materials, Zhejiang University, Hangzhou 310027, P. R. China}

\author{Man Cheng}
\affiliation{Zhejiang Province Key Laboratory of Quantum Technology and Device, Department of Physics, and State Key Laboratory of Silicon Materials, Zhejiang University, Hangzhou 310027, P. R. China}

\author{Zhou Liu}
\affiliation{State Key Laboratory for Mesoscopic Physics, School of Physics, Peking University, Beijing, 100871, P.R. China}
\affiliation{Beijing Key Laboratory for Magnetoelectric Materials and Devices, Beijing, 100871, P.R. China}

\author{Shijie Song}
\affiliation{Zhejiang Province Key Laboratory of Quantum Technology and Device, Department of Physics, and State Key Laboratory of Silicon Materials, Zhejiang University, Hangzhou 310027, P. R. China}

\author{Fanggui Wang}
\affiliation{State Key Laboratory for Mesoscopic Physics, School of Physics, Peking University, Beijing, 100871, P.R. China}
\affiliation{Beijing Key Laboratory for Magnetoelectric Materials and Devices, Beijing, 100871, P.R. China}

\author{Pimo He}
\affiliation{Zhejiang Province Key Laboratory of Quantum Technology and Device, Department of Physics, and State Key Laboratory of Silicon Materials, Zhejiang University, Hangzhou 310027, P. R. China}

\author{Guang-Han Cao}
\affiliation{Zhejiang Province Key Laboratory of Quantum Technology and Device, Department of Physics, and State Key Laboratory of Silicon Materials, Zhejiang University, Hangzhou 310027, P. R. China}
\affiliation{Collaborative Innovation Centre of Advanced Microstructures, Nanjing University, Nanjing 210093, P. R. China}

\author{Zhu-An Xu}
\affiliation{Zhejiang Province Key Laboratory of Quantum Technology and Device, Department of Physics, and State Key Laboratory of Silicon Materials, Zhejiang University, Hangzhou 310027, P. R. China}
\affiliation{Collaborative Innovation Centre of Advanced Microstructures, Nanjing University, Nanjing 210093, P. R. China}

\author{Yunhao Lu}
\email{luyh@zju.edu.cn}
\affiliation{Zhejiang Province Key Laboratory of Quantum Technology and Device, Department of Physics, and State Key Laboratory of Silicon Materials, Zhejiang University, Hangzhou 310027, P. R. China}

\author{Jinbo Yang}
\email{jbyang@pku.edu.cn}
\affiliation{State Key Laboratory for Mesoscopic Physics, School of Physics, Peking University, Beijing, 100871, P.R. China}
\affiliation{Beijing Key Laboratory for Magnetoelectric Materials and Devices, Beijing, 100871, P.R. China}

\author{Yi Zheng}
\email{phyzhengyi@zju.edu.cn}
\affiliation{Zhejiang Province Key Laboratory of Quantum Technology and Device, Department of Physics, and State Key Laboratory of Silicon Materials, Zhejiang University, Hangzhou 310027, P. R. China}
\affiliation{Collaborative Innovation Centre of Advanced Microstructures, Nanjing University, Nanjing 210093, P. R. China}

\date{\today}

\begin{abstract} The pivotal role of magnetic anisotropy in stabilising two-dimensional (2D) magnetism has been widely accepted \cite{CrI3_Nature_2017,CGT_Nature_2017,Science_Review_xiangzhang,FePS3_ising,NiPS_XY}, however, direct correlation between magnetic anisotropy and long-range magnetic ordering in the 2D limit is yet to be explored. Here, using angle- and temperature-dependent tunnelling magnetoresistance \cite{CrI3_pablo_2018, CrI3_xiaodongxu_2018,CrI3_verylarge,CrI3_10000TMR}, we report unprecedented metamagnetic phase transitions in atomically-thin CrOCl, triggered by magnetic easy-axis flipping instead of the conventional spin flop mechanism. Few-layer CrOCl tunnelling devices of various thicknesses consistently show an in-plane antiferromagnetic (AFM) ground state with the easy axis aligned along the Cr-O-Cr direction ($\mathbf{b}$-axis). Strikingly, with the presence of a magnetic field perpendicular to the easy-axis ($H\| \mathbf{c}$), magnetization of CrOCl does not follow the prevalent spin rotation and saturation pattern, but rather exhibits an easy-axis flipping from the in-plane to out-of-plane directions. Such magnetic anisotropy controlled metamagnetic phase transitions are manifested by a drastic upturn in tunnelling current, which shows anomalous shifts towards higher $H$ when temperature increases. By 2D mapping of tunnelling currents as a function of both temperature and $H$, we determine a unique ferrimagnetic state with a superstructure periodicity of five unit cells after the field-induced metamagnetic transitions. The feasibility to control 2D magnetism by manipulating magnetic anisotropy may open enormous opportunities in spin-based device applications.

\end{abstract} 

\maketitle


When magnetic anisotropy opens a gap in the low-energy modes of magnon excitations, local magnetic moments can form long range ordering in the 2D limit by various interaction mechanisms, such as direct and indirect exchanges, asymmetric Dzyaloshinskii-Moriya coupling, and Ruderman-Kittel-Kasuya-Yosida interaction via conduction electrons, etc \cite{Anderson1950,Kanamori1959,DM_inter,RKKY}. Prototypical examples are CrI$_{3}$ with a uniaxial out-of-plane magnetic anisotropy energy (MAE) and the Ising type of ferromagnetism (FM) \cite{CrI3_Nature_2017}, in contrast to Cr$_2$Ge$_2$Te$_6$ with an isotropic in-plane MAE and vanishing magnetic ordering when approaching the monolayer (ML) limit \cite{CGT_Nature_2017}. However, the recent gold rush in discovering divergent magnetic 2D materials \cite{CrBr3_2018,FGT_xiaodongxu_2018,FGT_yuanbozhang_2018,NiPS_XY,CrCl3_xiaodongxu,CrCl3_Pablo,CrCl3_Morpurgo,MnPS3_2020,CrPS4_ZY_2020} reveals that many van der Waals (vdW) magnets may have competing MAE axes, which induces distinctive metamagnetic transitions manifested as drastic magnetisation increases when external magnetic field ($H$) is applied along the easy axis of magnetization \cite{CrCl3_xiaodongxu,CrCl3_Pablo,CrCl3_Morpurgo,CrPS4_ZY_2020}. These abundant vdW metamagnets \cite{metamagnetism_review}, normally antiferromagnetic (AFM) in the bulk form, are only scantily explored but may open enormous opportunities in understanding and manipulating magnetism in the 2D limit, as exemplified by the discovery of magnetic anisotropy flipping controlled metamagnetic states in atomically thin CrOCl.

\begin{figure*}
\includegraphics[width=7 in]{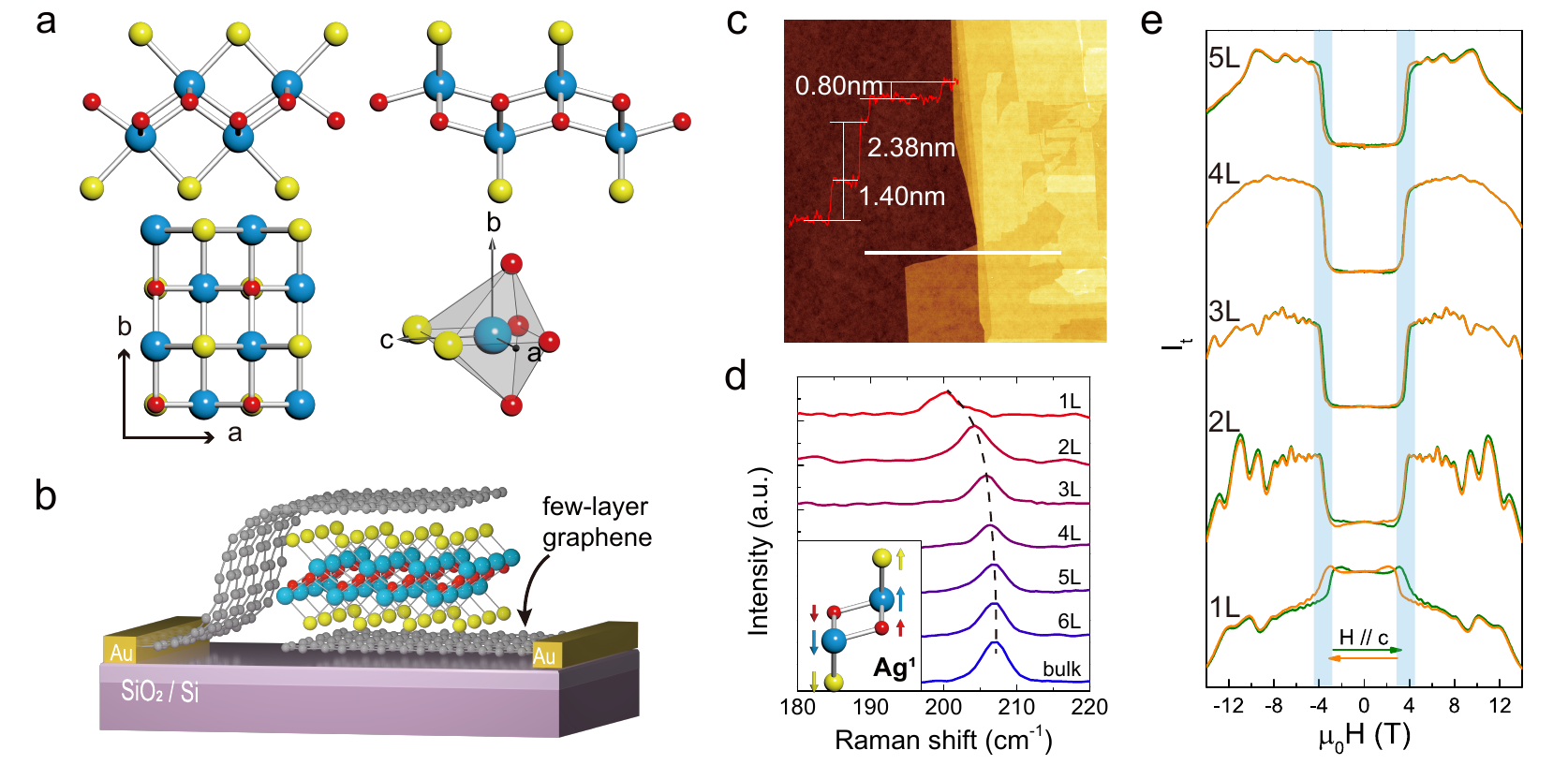}
\caption{\textbf{Few-layer tunnelling junctions of CrOCl.} \textbf{a}, Side- and top-view of the staggered square lattice of CrOCl consisting of CrO$_{4}$Cl$_{ 2}$ octahedrons, imposing cubic crystal field on 3\textit{d} orbitals in Cr$^{3+}$. \textbf{b}, Schematic cross-section of FLG/CrOCl/FLG tunnelling devices for TMR measurements. \textbf{c}, NCAFM image of a representative CrOCl flake, showing ML steps of $\sim0.8$ nm. Note that ML on SiO$_{2}$ exhibits an effective thickness of 1.40 nm. \textbf{d}, Fingerprinting Raman red shift in $A_{g}^{1}$ for determining CrOCl layer numbers.  \textbf{e}, $I_{t}$ vs $H$ characteristics of representative few-layer CrOCl tunnelling devices with various thicknesses from ML to quintuple layer. Note that high-frequency tunnelling peaks above 4 T are induced by the Shubnikov-de Haas oscillations of FLG electrodes.}
\label{fig1}
\end{figure*}


The bulk magnetism of CrOCl crystals have been systematically investigated in literatures \cite{CrOCl_neutron,CrOCl_prb_magnetoelastic}, which confirm an AFM ground state below N\'{e}el temperature $T_{N}$ of $\sim13.5$ K. Such an AFM ordering was interpreted as a \textbf{c}-easy axis type in which the magnetisation experiments show an unusual saturation behaviour with low magnetic moment per Cr atom of 0.6 $\mu_{B}$\cite{PRB2009-VOCl-magnetoelastic}. Recent $T$-dependent Raman scattering reveals the existence of an additional phase transition at $\sim 27$ K, possibly corresponding to the formation of an incommensurate magnetic superstructure from the paramagnetic high-$T$ phase\cite{PRB2009-VOCl-magnetoelastic,CrOCl_Raman}. These studies infer the critical importance of the unique quasi-1D staggered lattice of ML CrOCl, which is AA stacked out of plane to form the bulk. As shown in Figure \ref{fig1}a, ML CrOCl can be viewed as a staggered CrO double layer sandwiched by two halogen atomic layers. By occupying the same B-site, every two nearest neighbouring Cl and four O atoms form a distorted octahedron surrounding an A-site Cr, creating inequivalent upper and lower quasi-1D Cr chains along the Cl-Cr-Cl direction, \textit{i.e.} the \textbf{a} axis in Fig. \ref{fig1}a. In perpendicular to the 1D chains, Cr atoms are interconnected solely by O atoms, making the \textbf{b} axis significantly longer than \textbf{a} (lower panel of Fig. \ref{fig1}a).


Due to the wide semiconducting band gap of $\sim2.3$ eV, we fabricate vertical tunnelling junctions (TJs) to study the layer-dependent magnetism in CrOCl. Fig. \ref{fig1}b illustrates the tunnelling device structure, in which atomically-thin CrOCl channels are sandwiched by two few-layer graphene (FLG) electrodes. In brief, different-layer CrOCl flakes were first exfoliated on SiO$_2$/Si substrates, followed by two dry-transfer steps \cite{dry_transfer} to assemble a TJ device (see Methods for details). To minimize detrimental leakage current, we use narrow FLG flakes of several $\mu$m in width, and thus, confine the working channel areas to below $1.5$ $\mu$m$^{2}$. Typical tunnelling current ($I_{t}$) vs bias voltage ($V_\mathrm{bias}$) characteristics are shown in Supplementary Information (SI) Figure S1. After tunnelling experiments, the layer numbers of each device are determined by non-contact atomic force microscopy (NCAFM) using a ML step height of 0.8 nm (Fig. \ref{fig1}c). Equally important, Raman spectroscopy reveals that below six MLs, the $A_{g}^{1}$ phonon vibration mode shows monotonic red shift from the bulk value of $207.5$ cm$^{-1}$ to the ML wave number of $200.5$ cm$^{-1}$, allowing us to unambiguously confirm the layers of our devices in complementary with the NCAFM data (Fig. \ref{fig1}d and extended discussions in SI Note 1).


\begin{figure*}
\includegraphics[width=7 in]{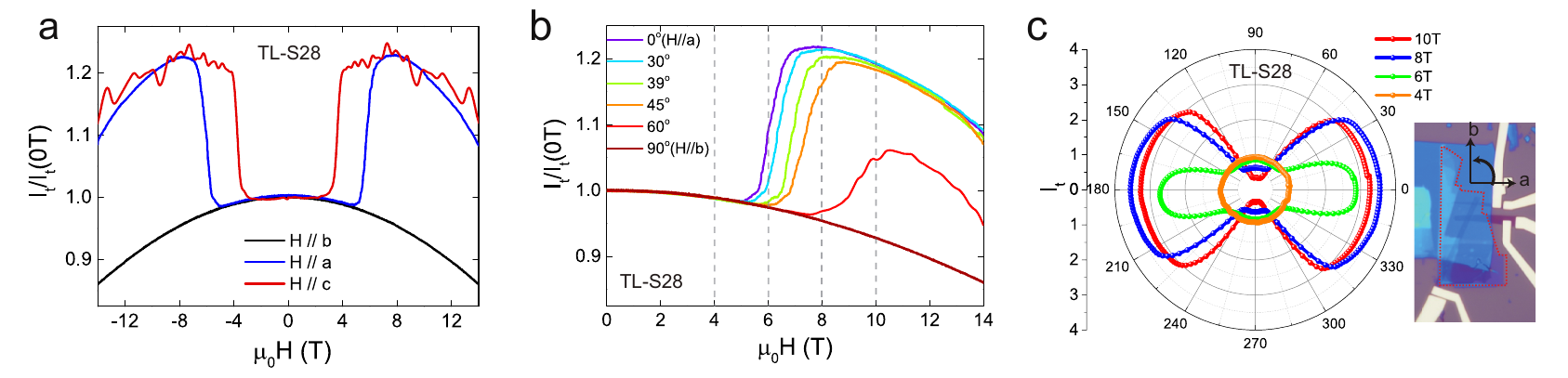}
\caption{\textbf{Angle-dependent tunnelling and magnetic field induced phase transitions in few-layer CrOCl devices.} 
\textbf{a}, Few-layer CrOCl devices consistently show steep $I_{t}$ upturns for both $H\| \mathbf{c}$ and $H\| \mathbf{a}$, which is contradicting with a uniaxial \textbf{c}-easy axis AFM. \textbf{b}, For in-plane $H$ rotation from the $\mathbf{a}$- to $\mathbf{b}$-axis, CrOCl TJ devices exhibit an inverse cosine scaling in the $I_{t}$ transition thresholds, suggesting Cr magnetic moments are rotated only by the effective magnetic field along the $\mathbf{a}$-axis. \textbf{c}, Polar plotting of $I_{t}$ as a function of in-plane $H$ rotation. The evolution of $I_{t}$ from isotropic circles in low $H$ to distinctive butterfly shapes in high $H$ manifests a field-induced magnetic phase transition.} 
\label{fig2}
\end{figure*}

Figure \ref{fig1}e summarizes layer-dependent $I_{t}$ as a function of out-of-plane $H$ at 1.6 K in few-layer CrOCl TJs with thicknesses from ML to quintuple layers, all consistently showing a drastic upturn in tunnelling current at around 3 T. At first glance, the results seem to suggest field induced spin-flop (SF) transitions for $H\| \mathbf{c}$, which are expected for a \textbf{c}-easy axis AFM ground state with negligible interlayer magnetic coupling. However, by rotating $H\| \mathbf{c}$ towards the $\mathbf{a}$ and $\mathbf{b}$ axes respectively, angle ($\theta$)-dependent $I_{t}$-$H$ curves unfold a distinct AFM magnetic ordering for few-layer CrOCl crystals. As shown in Figure \ref{fig2}a, abrupt $I_{t}$ upturns are observed for both $H\| \mathbf{c}$ and $H\| \mathbf{a}$, contradicting with the uniaxial \textbf{c}-easy axis AFM in the bulk. Because the $I_{t}$ upturn thresholds monotonically increase when $H$ rotates from the $\mathbf{c}$- to $\mathbf{a}$-axis (SI Figure S2), we can also exclude an easy-plane AFM configuration for few-layer CrOCl, which would show a maximum SF threshold when $H$ is aligned with the canting direction of Cr magnetic moments ($\theta \sim 30^{\circ}$; see extended discussions in SI Note 2). To get insight into the magnetic ground state of few-layer CrOCl TJs,  we further investigated the tunnelling behaviour with in-plane $H$ rotation. As shown in Figure \ref{fig2}b, when $H$ rotates from the $\mathbf{a}$- to $\mathbf{b}$-axis, there is an inverse cosine scaling in the $I_{t}$ transition thresholds, indicating that the effective magnetic field along the $\mathbf{a}$-axis is solely responsible for the magnetic phase transitions. Using polar plotting, it is clear to see that $I_{t}$, which is nearly isotropic in low $H$, is characterized by a distinctive butterfly-shaped two fold symmetry after the phase transitions (Figure \ref{fig2}c). 

\begin{figure*}
\includegraphics[width=4.5 in]{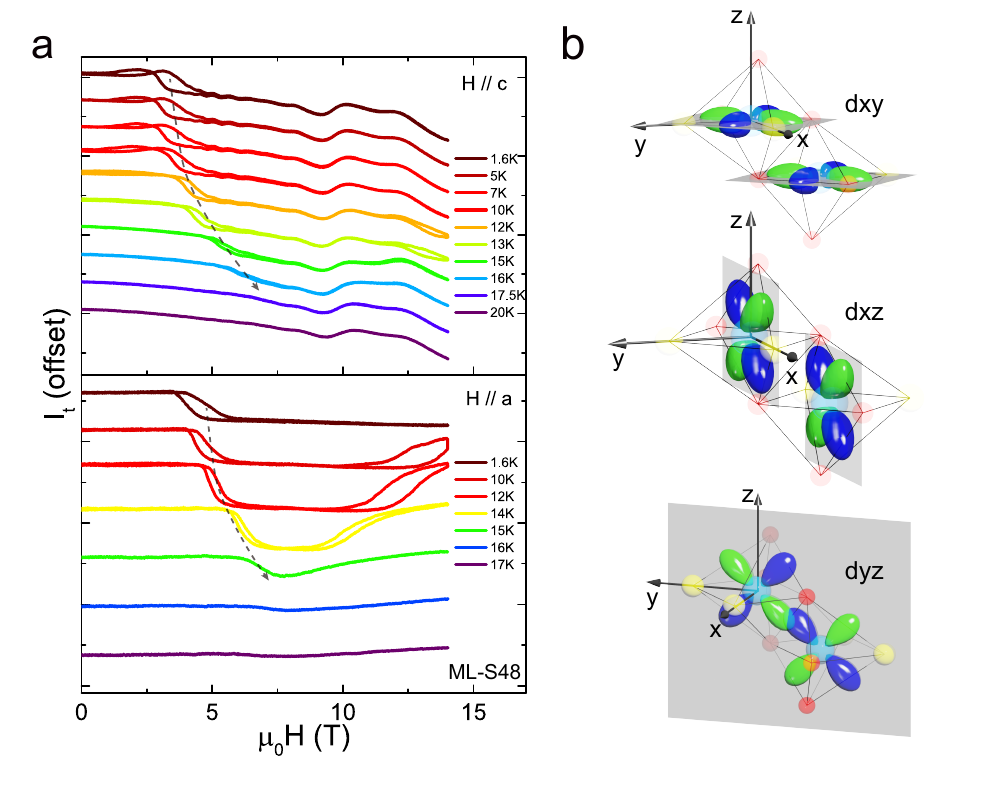}
\caption{\textbf{Temperature-dependent tunnelling in few-layer CrOCl devices.} 
\textbf{a}, $T$-dependent tunnelling of few-layer CrOCl devices reveal an opposite $T$ dependence to the trend of SF transition-controlled tunnelling, which should shift $I_{t}$ transition thresholds to lower $H$ when TJ devices are warmed up. \textbf{b}, Schematic of partial density of states of three occupied $t_{2g}$-$d$ orbitals in CrOCl. Due to the Cr$^{3+}$ valence state and an octahedron crystal field, $d_{zx}$, $d_{xy}$ and $d_{yz}$ orbitals of the $t_{2g}$ manifold are half-filled. Note that the conventional octahedron coordinates, in which $x$ and $y$ axes are defined by two Cr-Cl axes within the CrO$_{2}$Cl$_{ 2}$ basal plane, are used here for clarity.}
\label{fig3}
\end{figure*}

These self-contradicting experimental observations with the presumable bulk-like \textbf{c}-easy axis AFM ordering for few-layer CrOCl, however, can be consistently explained by an in-plane AFM ground state with its easy axis aligned along the \textbf{b} crystallographic direction. In this easy-axis configuration,  $H\| \mathbf{c}$ causes out-of-plane spin rotation in low fields, and eventually induces a magnetic phase transition which flips the easy magnetization direction from the \textbf{b}- to \textbf{c}-axis. Using $T$-dependent tunnelling, we can clearly see that the $I_{t}$ thresholds substantially increase when TJ devices are warmed up (Figure \ref{fig3}a). Such an anomalous $T$ dependence is opposite to the trend of SF transition-controlled tunnelling, which also can not explain the low-field hysteresis in all CrOCl TJs with different thicknesses (Figure \ref{fig3}a and Figure \ref{fig1}e).  

To appreciate such a field-induced MAE flipping mechanism, it is crucial to comprehend intralayer competition between in-plane exchange anisotropy (EA) and out-of-plane magnetocrystalline anisotropy (MCA), rooted in the unique staggered square lattice of CrOCl. In particularly, an in-plane AFM ordering in CrOCl requires a monoclinic lattice distortion by the rotation of the \textbf{a} axis within the $a-c$ plane \cite{CrOCl_prb_magnetoelastic}, rather different from the in-plane monoclinic transition in VOCl \cite{PRB2009-VOCl-magnetoelastic}. Despite sharing the same octahedron crystal field symmetry as CrOCl, V$^{3+}$ in VOCl has only two 3$d$ electrons occupying $d_{xy}$ and $d_{xz}$ orbitals respectively (Figure \ref{fig3}b and SI Note 3). By in-plane monoclinic distortion,  AFM ordering between neighbouring upper and lower V$^{3+}$ atoms along the compressed diagonal direction ($\frac{1}{2}\mathbf{a}+\frac{1}{2}\mathbf{b}$) is introduced by enhanced direct exchange interactions to avoid geometrical frustration associated with a rhombic unit cell (see SI Figure S3). Along the quasi-1D V-Cl-V atomic chains, direct exchange of filled $d_{xy}$ orbitals also enforce AFM coupling, leaving V$^{3+}$ atoms along the elongated ($\frac{1}{2}\mathbf{a}-\frac{1}{2}\mathbf{b}$) direction inevitably FM aligned.

However, once $d_{yz}$ is occupied in the case of CrOCl, in-plane lattice distortion becomes unfavourable because lattice compression along the ($\frac{1}{2}\mathbf{a}+\frac{1}{2}\mathbf{b}$) direction directly increases head-on overlapping between 3$d_{yz}$ orbitals (Figure \ref{fig3}b). Instead, CrOCl undergoes an out-of-plane lattice distortion by rotating the \textbf{a} axis towards \textbf{c}, which produces a \textbf{c}-axis monoclinic lattice of $90.06^\circ$ as well as zigzagging in the Cr-Cl-Cr atomic chains \cite{CrOCl_prb_magnetoelastic}. This unique type of monoclinic lattice distortion, which weakens direct exchange along the \textbf{a}-axis while leaving Cr-Cl-Cr superexchange interaction unchanged, makes the in-plane AFM ordering feasible in few-layer CrOCl when device encapsulations slightly reduce the monoclinic angle.  






\begin{figure*}
\includegraphics[width=7 in]{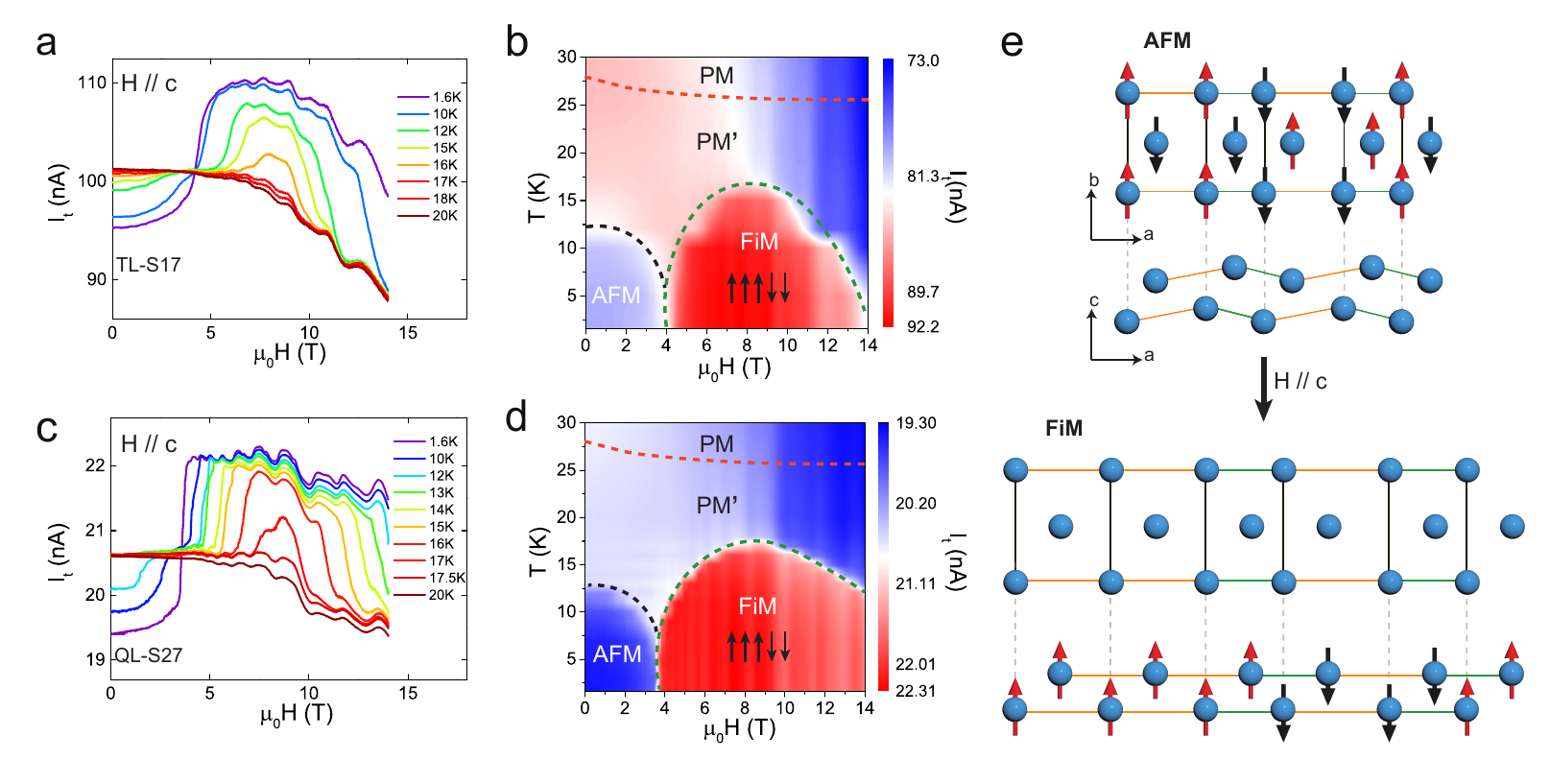}
\caption{\textbf{MAE flipping driven metamagnetism in few-layer CrOCl.} \textbf{a-d}, 2D mapping of $I_{t}$ as a function of both $T$ and $H$ for two representative TJ devices: a trilayer device TL-S17 (\textbf{a} and \textbf{b}) and a quadruple layer device QL-S27 (\textbf{c} and \textbf{d}). Few-layer CrOCl exhibits four distinctive magnetic phases, which are PM, PM$^{\prime}$, AFM and FiM, respectively. Note that the AFM dome and the much larger MAE-flipped dome are separated by the PM$^{\prime}$ phaseonly above 7 K. \textbf{e}, Top panel: Schematic of the in-plane AFM ground phase of few-layer CrOCl, which has a periodicity of four unit cells and quasi-1D FM chains along the \textbf{b}-axis. Bottom panel: The field-induced FiM state, which is an out-of-plane ferrimagnetic state with a five-fold periodicity. Note that there is a reentrant behaviour from the FiM to the PM$^{\prime}$ states above 8 T, when CrOCl devices are not at the base temperature.}
\label{fig4}
\end{figure*}

Using 2D mapping of $I_{t}$ as a function of both $T$ and $H$, it is illuminating to see that few-layer CrOCl exhibits four distinctive magnetic phases. As shown in Figure \ref{fig4}a-\ref{fig4}d for two representative devices (TL-S17 and QL-S27 respectively), few-layer CrOCl TJs in zero field make phase transitions at $\sim12.5$ K from an intemediate paramagnetic state (PM$^{\prime}$) to the AFM ground state, in good agreement with the bulk $T_{N}\sim13.5$ K. By applying $H\| \mathbf{c}$, the PM$^{\prime}$-to-AFM transitions are gradually suppressed, producing a magnetic dome structure which is typical for an AFM magnet. Above 4 T, there emerges a second magnetic dome with a distinct nature from a SF phase, i.e. canted AFM \cite{metamagnetism_review,CrCl3_xiaodongxu,CrCl3_Pablo,CrCl3_Morpurgo,CrPS4_ZY_2020}. As shown in Figure \ref{fig4}b and \ref{fig4}d, by fixing $T$ and sweeping $H$, well-defined phase boundaries between the AFM ground state and the MAE-flipped domain only exist below 7 K, above which the two magnetic domes are separated by the PM$^{\prime}$ phase. It is also conspicuous that, for fixed field of $H=8$ T, the MAE-flipped dome has a maximum phase transition $T$ of 17.5 K, which is significantly higher even when compared with the bulk $T_{N}$. Utilizing first-principle density functional theory (DFT) calculations (See detailed discussions in SI Note 4), we determine the magnetic ordering for the AFM ground state and the MAE-flipped phase, respectively. For the AFM ground state, we compare different in-plane magnetic configurations with zigzagging Cr-Cl-Cl atomic chains along the \textbf{a}-axis and varying \textbf{c}-axis monoclinic angles from $90^\circ$ to $90.06^\circ$. To meet the two criteria of minimized magnetic coupling energy and a \textbf{b}-easy axis AFM, we determine the a unique in-plane AFM state with a periodicity of four unit cells and quasi-1D FM chains along the \textbf{b}-axis. Based on this AFM configuration, the SF field is calculated to be exceeding 13 T, which can self-consistently explain the absence of SF phase transitions for  $H\| b$ (Fig. \ref{fig1}a and \ref{fig1}b). For the MAE-flipped phase, DFT suggests an out-of-plane ferrimagnetic (FiM) state with a five-fold periodicity (Bottom panel of Figure. \ref{fig4}e).  




Noticeably, few-layer CrOCl TJs also show abnormal downturn in $I_{t}$ for $H>8$ T, which can not be explained by field induced magnetic moment flipping within the ferrimagnetic state since increasing in magnetic moment alignments with $H$ generally enhance the tunnelling processes between two FLG electrodes. Such unusual tunnelling current decrease is due to the reentry of the FiM phase into the PM$^{\prime}$ phase. Indeed, as shown Figure \ref{fig4}a and Figure \ref{fig4}c, for both TL and QL CrOCl, the high-field $I_{t}$ for $H\| c$ approaches the tunnelling results of the PM$^{\prime}$ state. The trend becomes more conspicuous by warming up the samples to above 10 K, which clearly demonstrate that the high-field data at different $T$ set-points are converging to the PM$^{\prime}$ tunnelling curve of 20 K. It is also interesting to notice that 2D mapping of tunnelling magnetoresistance is able to resolve a definite phase boundary between the PM$^{\prime}$ and PM phases, by slow scanning of $T$ with fixed $H$ setpoints (SI Figure S4). This allows us to complete the magnetic phase diagram of few-layer CrOCl, as shown in Figure \ref{fig4}b and \ref{fig4}d.

Although atomically thin 2D materials in general exhibit distinctive physical properties from the bulk counterparts, e.g. interlayer coupling driven dimensional crossover in mechanical properties from ML to BL graphene and MoS$_{2}$ \cite{Dumitrica_prl11_GrRippling,ZhengY_prl15_MoS2-rippling}, the fascinating MAE flipping controlled magnetic ordering and divergent metamagnetism in few-layer CrOCl unfold the great promises of exploring 2D magnetism in a wide range of vdW layered AFM materials. Being isostructural compounds, it will be appealing to compare 2D magnetism in few-layer TiOCl, VOCl and FeOCl with CrOCl, considering that different numbers of $d$ electrons can drastically change the competition between in-plane EA and out-of-plane MCA. There may be other exotic magnetic mechanism in the 2D limit, such as spin-phonon coupling and magnetorestriction in these unique systems with staggered lattices.



\section*{Methods}
\textbf{Single Crystal Synthesis.} High-quality CrOCl single crystals were synthesised by chemical vapour transport of stoichiometrically mixed high-purity Cr$_{2}$O$_{3}$ and CrCl$_{3}$ powder (Alpha Esar, 99.9$\%$). The containing quartz tube was evacuated to a base pressure of 10$^{-2}$ Pa when flame-sealed before loading into a tube furnace. Single crystals were grown for 7 days in a temperature gradient of 1073 K to 1213 K. Then, the furnace was naturally cooled down to room temperature.


\textbf{Vibrating-Sample Magnetometer.} The bulk magnetisation of CrOCl single crystals were characterised using a vibrating sample magnetometer (VSM) module in a Quantum Design Physical Properties Measurement System (PPMS-9T). During the experiments, a CrOCl single crystal was vibrated sinusoidally, which causes an electrical signal proportional to the sample magnetisation in a stationary signal pick-up coil. The VSM results confirm the high quality of our single crystals, which are in excellent agreement with the literature reports \cite{CrOCl_prb_magnetoelastic}.

\textbf{Magnetic Torque.}
Magnetic torque measurements are based on a metal-cantilever capacitor geometry. During the torque experiment, a CrOCl single crystal was mounted on a cantilever made of Au foil, which has a thickness of 25 $\mu$m. The width of cantilever is 0.3 mm and the vacuum gap between cantilever and the bottom Au electrode is 0.5 mm. By applying an external magnetic field to the sample, torque exerted on the cantilever (${\tau}={\mu}_0m{\times}H$) are detected by measuring the capacitance signals, which are used to deduce the magnetizations of the sample.

\textbf{Tunnelling Device Fabrication.} Few-layer CrOCl flakes were exfoliated from single crystals in an argon-filled glove box to protect the samples from air degradation. Different layer numbers of CrOCl samples were first selected by optical contrast, which were cross-checked by complementary non-contact atomic force microscopy and Raman spectroscopy. During tunnelling junction preparation, top graphene electrodes, few-layer CrOCl and bottom graphene electrodes were picked up sequentially using PC on PDMS, followed by a final transferring step onto pre-patterned Cr/Au electrodes on Si/SiO$_2$ substrate. The resulting junction are free of interfacial contaminations, which is crucial for probing the intrinsic magnetism of few-layer CrOCl.   

\textbf{Tunnelling Magnetoresistance Measurements.}
Tunnelling measurements were performed in an Oxford-14 T cryostat. The system is equipped with a sample rotator for angle-dependent measurements with in-plane or out-of-plane magnetic field. The field- and $T$-dependent TMR and $I-V$ curves were measured using a Keithley 2400. 

\textbf{DFT Calculations.} To get the electronic structure of CrOCl, the Vienna \textit{ab initio} simulation package (VASP) \cite{VASP_Kresse_PRB93,VASP_Kresse_PRB96} is used, by the method of the projector augmented wave \cite{DFT_Blochl_PRB94} and the generalized gradient approximation (GGA-PBE) \cite{GGA_Perdew_PRL96}, in which the exchange-correlation potential was taken into account. The plane-wave cutoff energy is set at about 500 eV and the k-point sampling is performed by the Monkhorst-Pack scheme \cite{MPscheme_Monkhorst_PRB76}. The total energy is ensured to be converged within 1$\times$10$^{-6}$ eV per unit cell and the force threshold is 0.005 eV/\AA. For multilayer, the method of SCAN+rVV10 \cite{PRX-SCANrV10} is  adopted to describe the Van der Waals (vdW) interaction. The magnetic ground states of CrOCl are calculated using the method of PBE plus effective $U$\cite{LDAU-I_PRB1995}. For magnetic anisotropy energy (MAE) calculation, spin-orbit coupling was added as a perturbation term\cite{SOC1977}in VASP.

\textbf{Atomic Force Microscopy.} We use a multi-mode Park NX10 AFM to characterise the surface morphology and measure thin-film thicknesses of freshly exfoliated CrOCl crystals adn CrOCl TJ devices after tunnelling experiments. Few-layer microflakes were transferred onto a SiO$ _{2}$/Si substrate and baked \textit{in-situ} for a few minutes in order to enhance the contact between samples and the substrate.

\textbf{Raman Spectroscopy.} A Witec Alpha-300R Raman system with 532 nm laser excitation was used to perform Raman spectroscopy. The laser power was set to be below 5 mW to avoid excessive heating damage to the samples.

\section*{Data Availability} The authors declare that the main data supporting the findings of this study are available within the paper and its Supporting Information files. Extra data are available from the corresponding authors upon request.

\begin{acknowledgments} 
This work is supported by the National Key R\&D Program of the MOST of China (Grant Nos. 2016YFA0300204 and 2017YFA0303002), and the National Science Foundation of China (Grant Nos. 11790313 and 11574264), and Zhejiang Provincial Natural Science Foundation (D19A040001). Y.Z. acknowledges the funding support from the Fundamental Research Funds for the Central Universities.
\end{acknowledgments}

\section*{Author Contributions}
Y.Z. initiated and supervised the project. M.J.Z. and Q.F.H. synthesised and characterised COC crystals, fabricated COC devices and carried out all the measurements, assisted by M.C., Z. L. and F. W.. C.Q.H and Y.H.L. carried out the DFT calculations. M.J.Z., Q.F.H., C.Q.H., J.B.Y., Y.H.L. and Y.Z. analysed the data and wrote the paper with inputs from all authors. 

\section*{Competing financial interests:}
The authors declare no competing financial interests.

\section*{Data availability.}
The authors declare that the main data supporting the findings of this study are available within the paper and the Supplementary Information file. Extra data are available from the corresponding authors upon reasonable request. 

\end{document}